\documentclass{article}

\usepackage{graphicx}
\usepackage{amsmath}
\usepackage{aeguill}

\input{tcilatex}

\begin{document}

\title{Les structures fines de l'\'{e}lectromagn\'{e}tisme classique et de la
relativit\'{e} restreinte}
\author{Yves Pierseaux et Germain Rousseaux \\
ULB ypiersea@ulb.ac.be}
\maketitle

\begin{abstract}
L'un d'entre nous\ (Y. P.) a montr\'{e} l'existence d'une composante
longitudinale dans la propagation des ondes lumineuses sur la base de la
cin\'{e}matique sous-jacente \`{a} l'ellipse de Poincar\'{e}. \ Nous
montrons comment ce constat s'accorde avec la th\'{e}orie
\'{e}lectromagn\'{e}tique. Nous rappelons que l'autre d'entre nous soutient
l'existence d'une ''structure fine'' de l'electromagn\'{e}tisme \`{a} savoir
la coexistence de deux th\'{e}ories, l'une fond\'{e}e sur les champs
(Heaviside-Hertz) et l'autre sur les potentiels (Riemann-Lorenz).
L'existence de deux cin\'{e}matiques diff\'{e}rentes (''structure fine'' de
la relativit\'{e} restreinte: Poincar\'{e} ou Einstein) correspond \`{a} ces
deux formulations de l'electromagn\'{e}tisme classique.

Dans ce but, nous prouvons l'invariance relativiste de la d\'{e}composition
de Helmholtz du potentiel vecteur. Celle-ci se traduit par une compensation
g\'{e}n\'{e}ralis\'{e}e \`{a} toutes les directions de propagation, sur la
base de la tangente \`{a} l'ellipse de Poincar\'{e}, entre le potentiel
scalaire et la composante longitudinale du potentiel vecteur. L'adoption par
Poincar\'{e} de la condition de jauge de Lorenz (avec une composante
longitudinale et temporelle) est en contraste avec le photon einsteinien
avec uniquement des composantes transversales compatible avec le choix de la
condition de jauge de Coulomb ''compl\'{e}t\'{e}e''(jauge transverse).
\end{abstract}

\section*{Introduction}

\bigskip L'un d'entre nous a montr\'{e} que l'image par la transformation de
Lorentz (TL) d'un front d'onde plane (\`{a} deux dimensions spatiales: une
droite d'onde normale \`{a} la direction de propagation) \'{e}tait la
tangente \`{a} l'ellipse de Poincar\'{e}. D\`{e}s lors, si l'image
poincar\'{e}enne d'un front d'onde transversal n'est pas un front d'onde
transversal \cite{1}, cela \textit{semble} en contradiction avec la
th\'{e}orie \'{e}lectromagn\'{e}tique pour laquelle le caract\`{e}re
transversal des ondes \'{e}lectromagn\'{e}tiques planes ne fait aucun doute.
Or, comme l'ellipse avec sa tangente est enti\`{e}rement d\'{e}duite de la
TL, nous allons montrer que la diff\'{e}rence entre les deux points de vue,
Einstein ou Poincar\'{e}, (''la structure fine'' \cite{2}) ne se manifeste
pas au niveau des champs \'{e}lectromagn\'{e}tiques mais au niveau des
potentiels \'{e}lectromagn\'{e}tiques. Ceux qui sont par principe sceptiques
quant \`{a} l'existence d'une structure fine pourraient ainsi \^{e}tre
rassur\'{e}s puisque selon le point de vue de orthodoxe les potentiels ne ne
sont pas des \^{e}tres physiques. N'oublions cependant pas que, si du point
de vue pr\'{e}-relativiste le choix de jauge est consid\'{e}r\'{e} comme
arbitraire, le point de vue relativiste imposerait au contraire une jauge
bien d\'{e}termin\'{e}e qui devrait \^{e}tre covariante par la TL. Il
convient donc d'y regarder de plus pr\`{e}s.

Nous allons pr\'{e}ciser en quoi la structure fine de la relativit\'{e}
restreinte est reli\'{e}e \`{a} ce que l'on pourrait nommer la structure
fine de l'\'{e}lectromagn\'{e}tisme classique. En effet, il tout \`{a} fait
possible de formuler l'\'{e}lectromagn\'{e}tisme classique uniquement en
termes des potentiels \'{e}lectromagn\'{e}tiques d'apr\`{e}s les travaux de
Riemann et Lorenz en opposition avec la formulation orthodoxe de Heaviside
et de Hertz exprim\'{e}e uniquement en termes de champs
\'{e}lectromagn\'{e}tiques. Nous allons d\'{e}montrer la parfaite
compatibilit\'{e} entre la formulation de Riemann-Lorenz et la
cin\'{e}matique de Poincar\'{e}.

\section{La ''structure fine'' de l'\'{e}lectromagn\'{e}tisme classique}

La vision moderne de l'\'{e}lectromagn\'{e}tisme classique est issue des
travaux de H. Hertz et O. Heaviside qui prennent pour point de d\'{e}part de
la th\'{e}orie, les champs \'{e}lectrique et magn\'{e}tique qui sont
solutions des \'{e}quations dites ''de Maxwell'' o\`{u} interviennent les
densit\'{e}s de charge et de courant. R\'{e}soudre un probl\`{e}me d'\'{e}%
lectromagn\'{e}tisme revient donc \`{a} trouver les champs en fonctions des
sources. Cependant, en pratique, il est commode dans cette perspective
d'introduire des quantit\'{e}s math\'{e}matiques secondaires appel\'{e}es
potentiels qui sont d\'{e}finis de mani\`{e}re indirecte en fonction des
champs qui sont les quantit\'{e}s primaires observables. H.A.\ Lorentz
souligna que, de part cette d\'{e}finition indirecte, les potentiels sont ind%
\'{e}termin\'{e}s suivant ce que l'on appelle une ''transformations de
jauge'' qui laisse invariante les champs. Afin de lever cette ind\'{e}%
termination, il appara\^{i}t que l'on doit introduire dans les calculs en
fonction des potentiels une relation math\'{e}matique suppl\'{e}mentaire qui
fixe la valeur de ces potentiels: on choisit une ''jauge'' gr\^{a}ce \`{a}
une ''condition de jauge''. Historiquement, trois conditions de jauge \'{e}%
taient connues \`{a} l'\'{e}poque des travaux de Poincar\'{e} et d'Einstein 
\cite{3}\ \`{a} savoir principalement, la condition de jauge dite ''de
Coulomb'' (en fait due \`{a} Maxwell : $\nabla .\mathbf{A}=0$) et celle dite
''de Lorentz'' (en fait due \`{a} Kirchoff, Riemann, Lorenz, Lorentz et
Fitzgerald de mani\`{e}re ind\'{e}pendante : $\nabla .\mathbf{A}%
+1/c_{L}^{2}\partial _{t}V=0$ ; nous l'attribuons dans la suite \`{a} L.V.\
Lorenz car il f\^{u}t le premier \`{a} exprimer clairement que c'\'{e}tait
la ''condition'' de la propagation \`{a} c\'{e}l\'{e}rit\'{e} finie). On
introduit aussi la jauge dite ''transverse'' ou ''de rayonnement'' qui
consiste \`{a} postuler la condition de jauge de Coulomb et \`{a} annuler
simultan\'{e}ment le potentiel scalaire ($\nabla .\mathbf{A}=0$ et $V=0$).

La lecture des oeuvres de Maxwell pour un lecteur moderne est une source
d'\'{e}tonnement \cite{4}. En effet, contrairement \`{a} la vision actuelle
sur le sujet, Maxwell consid\`{e}re les potentiels comme des grandeurs
physiques \`{a} part enti\`{e}re : en particulier, le potentiel vecteur est
pour lui la ''quantit\'{e} fondamentale'' de l'\'{e}lectromagn\'{e}tisme
classique. Comment peut-on passer d'une vision o\`{u} les potentiels
semblent jouer un r\^{o}le pr\'{e}pond\'{e}rant \`{a} une vision o\`{u} ils
ne sont que des artifices math\'{e}matiques ? Il y a l\`a un fil d'Ariane
qu'il nous faut suivre \cite{5}. Nous allons montrer que la position
adopt\'{e}e par J.C.\ Maxwell \'{e}tait interm\'{e}diaire entre celle de
Riemann-Lorenz et celle qu'adoptera Heaviside et Hertz. Cette derni\`{e}re
sera adapt\'{e}e par H.A.\ Lorentz pour aboutir \`{a} la vision dominante
actuelle o\`{u} les potentiels ne sont pas exclus mais n'admettent pas
d'interpr\'{e}tation physique.

Avec la d\'{e}couverte par Hertz des ondes \'{e}lectromagn\'{e}tiques, il
s'est pos\'{e} la question de caract\'{e}riser l'orientation des champs \'{e}%
lectrique et magn\'{e}tique afin de tester une pr\'{e}diction essentielle de
Maxwell \`{a} savoir que la lumi\`{e}re consistait en une vibration \'{e}%
lectromagn\'{e}tique transverse \`{a} la direction de propagation d'une onde
plane \cite{6}. Or, Fitzgerald fit remarqu\'{e} qu'un dip\^{o}le \'{e}%
lectrique tel qu'utilis\'{e} par Hertz rayonne certes des ondes \'{e}%
lectromagn\'{e}tiques mais que celles-ci ne sont qu'approximativement planes
et transverses seulement loin de la source dans ce que l'on appelle d\'{e}%
sormais la zone de rayonnement \cite{7}. Proche de \ celui-ci, l'influence
du dip\^{o}le se r\'{e}sume \`{a} l'interaction Coulombienne obtenue par le
calcul en statique : la propagation est instantann\'{e}e. Hertz lui-m\^{e}me
a mis en \'{e}vidence exp\'{e}rimentalement l'existence de cette zone o\`{u}
la lumi\`{e}re semble se propager quasi-instantan\'{e}ment: en particulier,
loin de la source, on montre que la phase de l'onde hertzienne est d\'{e}cal%
\'{e}e d'un facteur $\pi $ par rapport \`{a} une onde qui se serait propag%
\'{e}e id\'{e}alement depuis cette m\^{e}me source \`{a} c\'{e}l\'{e}rit\'{e}
finie comme si la zone proche n'existait pas \cite{6}. De plus, on met aussi
en \'{e}vidence exp\'{e}rimentalement une zone dite interm\'{e}diaire o\`{u}
le rayonnement dip\^{o}laire \'{e}lectrique est tel qu'il existe une
composante radiale du champ \'{e}lectrique non-nulle qui se propage \`{a}
vitesse finie. Fitzgerald souligna justement que seule, la condition de
jauge de Lorenz, pouvait d\'{e}crire cette composante radiale d\'{e}pendante
du temps qui s'annulerait dans la jauge ''transverse'' (condition que
Maxwell avait utilis\'{e}e pour d\'{e}crire la propagation lumineuse; sinon
ce dernier utilisait la condition de jauge de Coulomb lorsqu'il n'y avait
pas d'ondes)\footnote{%
Selon Fitzgerald \cite{7}: \guillemotleft\ In most investigations on the
propagation of light attention has been concentrated on the transverse
nature of the vibration. Longitudinal motions have been relegated to the
case of pressural waves, and investigators have devoted themselves to
separating the two as much as possible... the existence of a longitudinal
component is mention only to show that it is very small and the motion is
mostly transverse. Now, the longitudinal component is no doubt generally
small except in the the immediate neighbourhood of a source; but it by no
means follows that, as a consequence, the actual direction of motion is
transverse at all points in a wave. In every complicated wave there are
points and often lines along which the transverse component vanishes, and at
all these places the small longitudinal component may be, and often is, of
great relative importance, so that the actual motion is largely in the
direction of wave propagation at these places. The simplest case is that of
a simple oscillator whose theory has been completely worked out by
Hertz...If the electric oscillator is parallel to z, we have for the
components of the vector potential :$A_{x}=0\quad A_{y}=0\quad A_{z}=A_{0}{%
\frac{{\cos (\omega t-kr)}}{r}}$ and the components of the electric force,
which are in general :${\frac{1}{{c^{2}}}}{\frac{{\partial \mathrm{E}_{%
\mathrm{x}}}}{{\partial \mathrm{t}}}}={\frac{{dJ}}{{dx}}}-\nabla
^{2}A_{x}\quad {\frac{1}{{c^{2}}}}{\frac{{\partial \mathrm{E}_{\mathrm{y}}}}{%
{\partial \mathrm{t}}}}={\frac{{dJ}}{{dy}}}-\nabla ^{2}A_{y}$ $\quad {\frac{1%
}{{c^{2}}}}{\frac{{\partial \mathrm{E}_{\mathrm{z}}}}{{\partial \mathrm{t}}}}%
={\frac{{dJ}}{{dz}}}-\nabla ^{2}A_{z}$ where : $J={\frac{{dA_{x}}}{{dx}}}+{%
\frac{{dA_{y}}}{{dy}}}+{\frac{{dA_{z}}}{{dz}}}$ become in this case : ${%
\frac{1}{{c^{2}}}}{\frac{{\partial \mathrm{E}_{\mathrm{x}}}}{{\partial 
\mathrm{t}}}}={\frac{{\partial ^{2}A_{z}}}{{\partial z\partial x}}}\quad {%
\frac{1}{{c^{2}}}}{\frac{{\partial \mathrm{E}_{\mathrm{y}}}}{{\partial 
\mathrm{t}}}}={\frac{{\partial ^{2}A_{z}}}{{\partial z\partial y}}}\quad $%
\par
${\frac{1}{{c^{2}}}}{\frac{{\partial \mathrm{E}_{\mathrm{z}}}}{{\partial 
\mathrm{t}}}}=-{\frac{{\partial ^{2}A_{z}}}{{\partial ^{2}x}}}-{\frac{{%
\partial ^{2}A_{z}}}{{\partial ^{2}y}}}$.
\par
It is particularly to be observed that $\partial _{t}E_{x}$ and $\partial
_{t}E_{y}$ arise entirely from $J$, which was dismissed by Maxwell as not
coming into consideration in cases of wave propagation on account of there
being no varying electrification. This is true as regards \textit{propagation%
}, but not all as regards origination. In all cases of origination we have
to do with conduction, or its equivalent convection, and in most such cases
we have changing electrification which brings in the $J$ term.%
\guillemotright }.

Peu apr\`{e}s les travaux du physicien G.F. Fitzgerald et
ind\'{e}pendamment, le math\'{e}maticien T. Levi-Civita publia un article
dans lequel il se donnait pour but selon ses termes de ''r\'{e}duire'' la
th\'{e}orie de Helmholtz \`{a} celle de Hertz \cite{8}. En fait la
th\'{e}orie de Helmholtz dont parle Levi-Civita consiste en la th\'{e}orie
de Riemann-Lorenz\footnote{%
A la fin du XIX\`{e}me si\`{e}cle, la th\'{e}orie de Maxwell n'est pas
universellement adopt\'{e}e et plusieurs autres th\'{e}ories sont en
comp\'{e}tition. H. Poincar\'{e} fit un remarquable recensement des
th\'{e}ories existantes dans son livre ''Electricit\'{e} et Optique'' \cite
{9}. Il distinguait la th\'{e}orie de Maxwell, la th\'{e}orie de Helmholtz,
la th\'{e}orie de Hertz et la th\'{e}orie de Lorentz.}, \`{a} savoir que les
potentiels sont solutions d'une \'{e}quation de D'Alembert avec un terme
source proportionnel \`{a} la densit\'{e} de charge ou de courant : 
\begin{equation}
\nabla ^{2}V-{\frac{1}{{c_{L}^{2}}}}{\frac{{\partial ^{2}V}}{{\partial t^{2}}%
}}=-{\frac{\rho }{{\varepsilon _{0}}}}
\end{equation}
\begin{equation}
\nabla ^{2}\mathbf{A}-{\frac{1}{{c_{L}^{2}}}}{\frac{{\partial ^{2}\mathbf{A}}%
}{{\partial t^{2}}}}=-\mu _{0}\mathbf{j}
\end{equation}
La th\'{e}orie de Helmholtz est une variante de la th\'{e}orie de
Riemann-Lorenz en ce sens qu'elle pr\'{e}voit contrairement \`{a} la
th\'{e}orie de Maxwell l'existence d'ondes longitudinales de potentiel
vecteur mais qui se propagent \`{a} une vitesse diff\'{e}rente des ondes
transverses de potentiel vecteur de mani\`{e}re analogue \`{a} la
propagation des ondes \'{e}lastiques \cite{6}. Cependant, le potentiel
scalaire se propage de mani\`{e}re instantan\'{e}e chez Helmholtz comme chez
Maxwell. Dans la th\'{e}orie de Riemann-Lorenz, il existe une propagation
longitudinale des potentiels scalaire et vecteur \`{a} la m\^{e}me
c\'{e}l\'{e}rit\'{e} (de la lumi\`{e}re) que celle des ondes transverses en
potentiel vecteur \cite{10}. Levi-Civita f\^{u}t le premier \`{a}
d\'{e}montrer math\'{e}matiquement que les \'{e}quations de propagation pour
les potentiels auxquelles on ajoute la conservation de la charge ($\nabla .%
\mathbf{j}+\partial _{t}\rho =0$) permettent de retrouver l'\'{e}quation de
Lorenz ($\nabla .\mathbf{A}+1/c_{L}^{2}\partial _{t}V=0$) ainsi que les
\'{e}quations de Heaviside-Hertz en termes des champs \cite{8} \& \cite{11}
: 
\begin{equation*}
\nabla .\mathbf{B}=0
\end{equation*}
\begin{equation}
\partial _{t}\mathbf{B}=\mathrm{\ -}\nabla \times \mathbf{E}
\end{equation}
\begin{equation*}
\nabla .\mathbf{E}={\frac{\rho }{{\varepsilon _{0}}}}
\end{equation*}
\begin{equation}
\nabla \times \mathbf{B}=\mu _{\mathrm{0}}\mathbf{j}+{\frac{1}{{c_{L}^{2}}}}%
\partial _{t}\mathbf{E}
\end{equation}
On peut aussi postuler l'\'{e}quation de Lorenz et retrouver l'\'{e}quation
de conservation de la charge. Donc, la th\'{e}orie de Riemann-Lorenz est
math\'{e}matiquement plus fondamentale que celle de Heaviside-Hertz qui en
d\'{e}coule \cite{5}. Quant \`{a} Lorentz, il postulait la formulation de
Heaviside-Hertz, introduisait les potentiels \`{a} partir des champs $%
\mathbf{B}=\nabla \times \mathbf{A}$ et $\mathbf{E}=\mathrm{\ -}\partial _{%
\mathrm{t}}\mathbf{A}\mathrm{\ -}\nabla V$ et constatait qu'ils \'{e}taient
ind\'{e}termin\'{e}s selon les transformations de jauge \cite{4}.

A ce niveau de notre enqu\^{e}te historique, il est essentiel de se poser la
question suivante : la formulation de Riemann-Lorenz pourrait-elle \^{e}tre
une alternative \`{a} celle de Lorentz ? En effet, si l'on fait
l'hypoth\`{e}se que les potentiels sont des quantit\'{e}s primaires dont
d\'{e}rivent les champs en tant que quantit\'{e}s secondaires alors il faut
pouvoir d\'{e}finir les potentiels de mani\`{e}re ind\'{e}pendante des
champs et expliquer pourquoi la condition de jauge de Lorenz est une
n\'{e}cessit\'{e} (une contrainte) et pas un choix commode pour les calculs.
De plus, que devient la condition de jauge de Coulomb (ou transverse)
puisque la formulation de Riemann-Lorenz s\'{e}lectionne d'embl\'{e}e celle
de Lorenz alors que la formulation de Lorentz les met sur un pied
d'\'{e}galit\'{e} en tant que "fixatrices de jauge"? Pourquoi la formulation
de Lorentz a-t-elle \'{e}t\'{e} adopt\'{e}e plut\^{o}t que celle de
Riemann-Lorenz ?

L'ensemble de ces questions cristallise ce que nous appellerons
d\'{e}sormais la "structure fine" de l'\'{e}lectromagn\'{e}tisme classique.

Quelle signification physique peut-on attribuer \`{a} l'\'{e}quation de
Lorenz qui semble jouer un r\^{o}le pr\'{e}pond\'{e}rant dans la formulation
de Riemann-Lorenz ? Tout d'abord, on peut remarquer qu'elle a la forme d'une
\'{e}quation de continuit\'{e} : divergence d'un flux plus d\'{e}riv\'{e}e
temporelle d'une densit\'{e} \'{e}gale \`{a} z\'{e}ro. Par ailleurs, Riemann
a fait remarqu\'{e}, il y a bien longtemps, la ressemblance entre cette
\'{e}quation et l'\'{e}quation de continuit\'{e} d'un fluide ($\rho \nabla .%
\mathbf{u}+D\rho /Dt=0$). Rousseaux \cite{10} a r\'{e}cemment
pr\'{e}cis\'{e} la pens\'{e}e de Riemann en montrant que l'\'{e}quation de
continuit\'{e} hydrodynamique pouvait s'\'{e}crire sous une forme
strictement analogue \`{a} l'\'{e}quation de Lorenz en consid\'{e}rant des
ondes acoustiques ($\nabla .\delta \mathbf{u}+1/c_{S}^{2}\partial
_{t}(\delta p/\rho _{0})=0$). En hydrodynamique, la propagation de la
perturbation de vitesse est indissociable de celle de la perturbation de
pression : soit l'\'{e}coulement est incompressible et il n'y a pas d'ondes
acoustiques, soit l'\'{e}coulement est compressible et il y a \`{a} la fois
des ondes de pression et de vitesse. Par analogie avec la m\'{e}canique des
fluides, si les potentiels se propagent selon une \'{e}quation de Riemann
alors ils doivent ob\'{e}ir \`{a} une \'{e}quation de continuit\'{e}
\'{e}lectromagn\'{e}tique \`{a} savoir la ''contrainte de Lorenz'' ainsi que
nous la d\'{e}signerons d\'{e}sormais.\newline

Quelles significations physiques peut-on attribuer aux potentiels ? Maxwell
appelait le potentiel vecteur soit l'intensit\'{e} \'{e}lectrotonique, soit
la quantit\'{e} de mouvement \'{e}lectrocin\'{e}tique, soit la quantit\'{e}
de mouvement \'{e}lectromagn\'{e}tique. Clairement, il l'identifiait \`{a}
une impulsion g\'{e}n\'{e}ralis\'{e}e au sens de la M\'{e}canique Analytique
de Lagrange \cite{4}.

En effet, selon Maxwell :

\begin{quotation}
\guillemotleft\ The conception of such a quantity, on the changes of which,
and not on its absolute magnitude, the induction currents depends, occurred
to Faraday at an early stage of his researches. He observed that the
secondary circuit, when at rest in an electromagnetic field which remains of
constant intensity, does not show any electrical effect, whereas, if the
same state of the field had been suddenly produced, there would have been a
current. Again, if the primary circuit is removed from the field, or the
magnetic forces abolished, there is a current of the opposite kind. He
therefore recognised in the secondary circuit, when in the electromagnetic
field, a ``peculiar electrical condition of matter'\ to which he gave the
name of Electrotonic State.\guillemotright
\end{quotation}

Selon William Whewell (un des professeurs de Maxwell), l'\'{e}tat
\'{e}lectro-tonique de Faraday se traduit par l'existence d'une quantit\'{e}
de mouvement dans le milieu. La r\'{e}sistance du milieu \`{a} la formation
d'un courant est analogue \`{a} l'inertie qui s'oppose \`{a} la mise en
mouvement d'un objet mat\'{e}riel: \newline

\begin{center}
$\mathbf{F}={\frac{{d\mathbf{p}}}{{dt}}}$ $\Leftrightarrow $ Capacit\'{e}
\`{a} produire un courant = variation temporelle de l'\'{e}tat
\'{e}lectro-tonique

Inductance d'un circuit $\Leftrightarrow $ Inertie d'une masse
\end{center}

Maxwell identifie le potentiel vecteur moderne comme \'{e}tant
l'intensit\'{e} \'{e}lectro-tonique et dont la d\'{e}riv\'{e}e temporelle,
en l'occurrence le champ \'{e}lectrique, produit une force
\'{e}lectro-motrice :

\begin{center}
$\mathbf{F} = {\frac{{d\mathbf{p}} }{{dt}}}$ $\Leftrightarrow $ $\mathbf{E}
= - {\frac{{d\mathbf{A}} }{{dt}}}$
\end{center}

La loi empirique de Lenz (1834) sur l'induction explique le signe
n\'{e}gatif.

Selon Maxwell \cite{4} :

\begin{quotation}
\guillemotleft\ The Electrokinetic momentum at a point represents in
direction and magnitude the time-integral of the electromotive intensity
which a particle placed at this point would experience if the currents were
suddenly stopped. Let Ax, Ay, Az represent the components of the
electromagnetic momentum at any point of the field, due to any system of
magnets or currents. Then Ax is the total impulse of the electromotive force
in the direction of x that would be generated by the removal of these
magnets or currents from the field, that is, if Ex be the electromotive
force at any instant during the removal of the system : $A_{x}=\int {E_{x}dt}
$. Hence the part of the electromotive force which depends on the motion of
magnets or currents in the field, or their alteration of intensity, is : $%
E_{x}=-{\frac{{\partial A_{x}}}{{\partial t}}}$...If there is no motion or
change of strength of currents or magnets in the field, the electromotive
force is entirely due to variation of electric potential, and we shall have
: $E_{x}=-{\frac{{\partial V}}{{\partial x}}}$ ...\guillemotright
\end{quotation}

Le potentiel vecteur est une impulsion \'{e}lectromagn\'{e}tique c.\`{a}.d.
une variation de quantit\'{e} de mouvement \'{e}lectromagn\'{e}tique selon
la d\'{e}finition m\'{e}canique d'une impulsion. Or, il est n\'{e}cessaire
de se donner un r\'{e}f\'{e}rentiel pour d\'{e}finir une impulsion
m\'{e}canique ce qui se traduit par l'existence d'une constante de
r\'{e}f\'{e}rence pour le potentiel vecteur. D'une mani\`{e}re moderne, on
peut donc donner la d\'{e}finition suivante : le potentiel vecteur en un
point M est l'impulsion qu'un op\'{e}rateur ext\'{e}rieur doit fournir
m\'{e}caniquement \`{a} une charge unit\'{e} pour l'amener de l'infini,
o\`{u} par convention celui-ci est nul, jusqu'au point M.

Concernant le potentiel scalaire, il suffit de remplacer le mot impulsion
par \'{e}nergie dans la d\'{e}finition pr\'{e}c\'{e}dente en se rappelant
que d'apr\`{e}s Maxwell : \guillemotleft\ Potential, in electrical science,
has the same relation to Electricity that Pressure, in Hydrostatics, has to
Fluid\UNICODE{0x2026} Electricity and Fluids all tend to pass from one place
to another if the Potential, Pressure is greater in the first place than in
the second. \guillemotright \newline

Pourquoi Maxwell n'a-t-il pas propos\'{e} la th\'{e}orie de Riemann-Lorenz
alors qu'il connaissait leur travaux ? La r\'{e}ponse se trouve dans une
note \'{e}crite par Maxwell sur la th\'{e}orie \'{e}lectromagn\'{e}tique de
la lumi\`{e}re \cite{4} :

\begin{quotation}
\guillemotleft\ From the assumption of both these papers we may draw the
conclusions, first, that action and reaction are not always equal and
opposite, and second, that apparatus may be constructed to generate any
amount of work from its own resources. For let two oppositely electrified
bodies A and B travel along the line joining them with equal velocities in
the direction AB, then if either the potential or the attraction of the
bodies at a given time is that due to their position at some former time (as
these authors suppose), B, the foremost body, will attract A forwards more
than B attracts A backwards. Now let A and B be kept asunder by a rigid rod.
The combined system, if set in motion in the direction AB, will pull in that
direction with a force which may either continually augment the velocity, or
may be used as an inexhaustible source of energy. \guillemotright
\end{quotation}

En clair, Maxwell a rejet\'{e} les cons\'{e}quences ''relativistes'' de la
th\'{e}orie de Riemann-Lorenz en ce sens que la troisi\`{e}me loi de Newton
n'est plus valide en relativit\'{e} restreinte car elle pr\'{e}suppose la
simultan\'{e}it\'{e} absolue de l'action et de la r\'{e}action ce qui est
incompatible avec le principe de relativit\'{e} comme l'a d\'{e}montr\'{e}
Henri Poincar\'{e} (\cite{12} \& \cite{13}): deux \'{e}v\'{e}nements
\'{e}tant observ\'{e}s dans un r\'{e}f\'{e}rentiel qui semblent
simultan\'{e}s, ne sont plus simultan\'{e}s \'{e}tant observ\'{e}s dans un
autre r\'{e}f\'{e}rentiel. En effet, on peut lire par exemple dans Science
et M\'{e}thode :

\begin{quotation}
''Voyons ce que devient, dans la th\'{e}orie de Lorentz, le principe de
l'\'{e}galit\'{e} de l'action et de la r\'{e}action. Voil\`{a} un
\'{e}lectron A qui entre en mouvement pour une cause quelconque?; il produit
une perturbation dans l'\'{e}ther?; au bout d'un certain temps, cette
perturbation atteint un autre \'{e}lectron B, qui sera d\'{e}rang\'{e} de sa
position d'\'{e}quilibre. Dans ces conditions, il ne peut y avoir
\'{e}galit\'{e} entre l'action et la r\'{e}action, au moins si l'on ne
consid\`{e}re pas l'\'{e}ther, mais seulement les \'{e}lectrons qui sont
seuls observables, puisque notre Nature est form\'{e}e d'\'{e}lectrons. En
effet, c'est l'\'{e}lectron A qui a d\'{e}rang\'{e} l'\'{e}lectron B?; alors
m\^{e}me que l'\'{e}lectron B r\'{e}agirait sur A, cette r\'{e}action
pourrait \^{e}tre \'{e}gale \`{a} l'action, mais elle ne saurait, en aucun
cas, \^{e}tre simultan\'{e}e, puisque l'\'{e}lectron B ne pourrait entrer en
mouvement qu'apr\`{e}s un certain temps, n\'{e}cessaire pour la
propagation.''
\end{quotation}

Maintenant, la th\'{e}orie propos\'{e}e par Maxwell reposait sur son
utilisation exclusive de l'\'{e}quation de Coulomb (ou la jauge transverse
pour les ondes) qui imposait une propagation instantan\'{e}e du potentiel
scalaire qui, de notre point de vue, avait deux avantages absolument
cruciaux pour Maxwell : le premier \'{e}tait que cette propagation
instantan\'{e}e garantissait la simultan\'{e}it\'{e} absolue et donc la
validit\'{e} de la loi de Coulomb ainsi que la troisi\`{e}me loi de Newton ;
le second \'{e}tait que le potentiel vecteur (donc les champs) \'{e}tait
transverse pour une onde \'{e}lectromagn\'{e}tique plane en accord avec la
th\'{e}orie optique de la lumi\`{e}re qui avait d\'{e}montr\'{e} le
caract\`{e}re transverse de la lumi\`{e}re gr\^{a}ce aux exp\'{e}riences de
polarisation de Fresnel.

Le premier avantage a \'{e}t\'{e} rejet\'{e} par Poincar\'{e} comme nous
l'avons rappel\'{e} plus haut. Nous insistons particuli\`{e}rement sur la
filiation entre l'exp\'{e}rience de pens\'{e}e de Maxwell \`{a} propos du
principe de r\'{e}action et les raisonnements ult\'{e}rieurs de Poincar\'{e}
qui pouss\`{e}rent ce dernier \`{a} adopter la formulation de Riemann-Lorenz
o\`{u} le potentiel scalaire se propage \`{a} vitesse finie en contradiction
avec la troisi\`{e}me loi de Newton. Par ailleurs, le deuxi\`{e}me article
de Levi-Civita f\^{u}t publi\'{e} en fran\c{c}ais dans les annales de la
facult\'{e} des sciences de Toulouse en 1902 apr\`{e}s la parution de la
seconde \'{e}dition d'Electricit\'{e} et Optique en 1901 que Poincar\'{e}
n'aurait pas manqu\'{e} de recencer. De plus, H.A. Lorentz cita d\`{e}s 1903
le premier article en italien de L\'{e}vi-Civita dans une publication en
anglais \cite{14}. Il est tr\`{e}s probable que Poincar\'{e} ait eu vent des
travaux de L\'{e}vi-Civita soit directement soit indirectement via Lorentz
apr\`{e}s 1902. Ainsi, il n'est pas \'{e}tonnant de constater que
Poincar\'{e} ait adopt\'{e} la formulation de Riemann-Lorenz dans son
article de 1906 sur la dynamique de l'\'{e}lectron. Celle-ci joua un
tr\`{e}s grand r\^{o}le car il p\^{u}t d\'{e}montrer l'existence d'un
quadri-vecteur potentiel ainsi que la covariance de la contrainte de Lorenz
\`{a} partir de celle de la conservation de la charge en ayant postul\'{e}
les transformations de l'espace et du temps qu'il attribua \`{a} Lorentz en
utilisant l'invariance de la charge totale. En juillet 1912, quelques jours
avant sa mort, Poincar\'{e} exprimera de mani\`{e}re indiscutable son
adoption de la r\'{e}duction de Levi-Civita \cite{19}:

\begin{quotation}
''Les \'{e}quations de la seconde colonne (la formulation de Heaviside-Hertz
: note des auteurs) \'{e}tant des cons\'{e}quences de celles de la
premi\`{e}re (la formulation de Riemann-Lorenz : note des auteurs)''
\end{quotation}

Le second avantage a \'{e}t\'{e} rejet\'{e} par l'un des auteurs (G.R.) en
montrant que l'utilisation de l'\'{e}quation de Coulomb \'{e}tait
confin\'{e}e \`{a} la limite galil\'{e}enne dite \guillemotleft\
magn\'{e}tique \guillemotright\ de l'\'{e}lectromagn\'{e}tisme classique due
\`{a} L\'{e}vy-Leblond \& Le Bellac \cite{10} \& \cite{17} et ne pouvait
donc pas d\'{e}crire la propagation de la lumi\`{e}re car elle est
l'approximation de la contrainte de Lorenz qui est covariante selon les
transformations de Poincar\'{e}-Lorentz et dont on prend la limite. Cette
derni\`{e}re devient covariante galil\'{e}enne dans la limite dite
\guillemotleft\ \'{e}lectrique \guillemotright .

\section{La ''structure fine'' de la relativit\'{e} restreinte}

Contrairement \`{a} Poincar\'{e}, Einstein n'\'{e}voque jamais les
potentiels dans ses articles sur la relativit\'{e}. Il semble donc difficile
d'admettre que ces derniers jouent un r\^{o}le quelconque (et donc a
fortiori un r\^{o}le essentiel dans la cin\'{e}matique einsteinienne).
Cependant, l'un d'entre nous (Y.P.) a montr\'{e} que \cite{1}, si la
d\'{e}finition de la propagation longitudinale de l'onde plane ne posait
aucun probl\`{e}me, la g\'{e}n\'{e}ralisation \`{a} toute direction de
propagation posait par contre un probl\`{e}me s\'{e}rieux (exactement comme
la composition de deux vitesses dans des directions diff\'{e}rentes dans la
pr\'{e}cession de Thomas). Nous avons ainsi montr\'{e} que la d\'{e}finition
einsteinienne de l'onde plane imposait la transversalit\'{e} de l'onde par
rapport \`{a} la direction de propagation dans les deux syst\`{e}mes K et K'
(objet et image). Le front image einsteinien, obtenu par la TL d'un front
perpendiculaire \`{a} la direction de propagation dans K, selon ses propres
termes, est \textit{aussi} perpendiculaire \`{a} la direction de propagation
dans K'($\varphi $ \'{e}tant l'angle dans le syst\`{e}me de la source):

\begin{quotation}
En appelant $\varphi ^{\prime }$ l'angle form\'{e} par \textbf{la normale de
l'onde} (direction radiale) dans le syst\`{e}me en mouvement et la
''direction du mouvement''...\footnote{%
En fait dans le texte original, Einstein s'est tromp\'{e} et \`{a} \'{e}crit
\`{a} la place de ''la direction du mouvement'', ''la ligne
observateur-source''. Il a corrig\'{e} sur son exemplaire personnel (voir
Balibar, Einstein, Oeuvres choisies, Relativit\'{e}s I, page 49 note 56).} 
\cite[paragraphe 7]{21}
\end{quotation}

\bigskip Nous avons montr\'{e} (\`{a} deux dimensions spatiales et en pla\c{c%
}ant une source \`{a} l'infini au repos dans K ) que l'image par la TL du
front plan (une droite d'onde) dans K' \'{e}tait la tangente \`{a} l'ellipse
allong\'{e}e de Poincar\'{e} \cite{22} qui ne forme \'{e}videmment pas un
angle droit avec la direction de propagation dans K'. L'interpr\'{e}tation
physique de la tangente \`{a} l'ellipse est imm\'{e}diate: un ensemble
d'\'{e}v\'{e}nements simultan\'{e}s dans K ne correspond pas, par la TL,
\`{a} un ensemble d'\'{e}v\'{e}nenements simultan\'{e}s dans K'. La \textit{%
double transversalit\'{e}} einsteinienne (ou la double simultan\'{e}it\'{e}
einsteinienne) consiste alors \`{a} annuler la composante longitudinale d'un
vecteur \textbf{d'} situ\'{e} sur cette tangente \cite{1}. Rappelons que ce
vecteur\textbf{\ d'} est l'image par la TL d'un vecteur \textbf{d}
transverse \`{a} la direction de propagation (avec \textbf{k} pour vecteur
d'onde) dans le syst\`{e}me de la source K. Nous avons \'{e}galement
montr\'{e} que la composante transversale de ce vecteur \textbf{d'}
\'{e}tait invariante: $d=d_{\perp }=d_{\perp }^{\prime }$

Toutefois cette ''double'' transversalit\'{e} peut a priori se rapporter aux
champs \textbf{E }et \textbf{E'} ou au potentiel vecteur \textbf{A} et 
\textbf{A'}.

\subsection{La cin\'{e}matique d'Einstein et la condition de jauge
''transverse'' (Coulomb compl\'{e}t\'{e}e)}

Montrons maintenant que l'\'{e}criture einsteinienne de l'onde plane ($%
\theta \neq \theta ^{\prime }\neq 0$) implique non seulement la\ double
transversalit\'{e} au niveau des champs \textit{mais aussi et surtout} au
niveau du \textbf{potentiel vecteur }\cite{1}. Il y a deux possibilit\'{e}s:
ou bien le vecteur math\'{e}matique \textbf{d} sur le front d'onde objet
repr\'{e}sente soit le vecteur champ \'{e}lectrique \textbf{E }ou le champ
magn\'{e}tique \textbf{B}, soit le potentiel vecteur \textbf{A}.

En se basant sur les TL du champ \'{e}lectromagn\'{e}tique \'{e}crites par
Einstein aux \S 6 et 8, on obtient la transformation relativiste de la
composante transversale du champ \'{e}lectrique \cite{21}:

\begin{equation}
\mathbf{E}_{\perp }^{\prime }=\gamma (\mathbf{E}_{\perp }+\mathbf{v\times B)}
\end{equation}

Il n'est pas restrictif de travailler \`{a} deux dimensions spatiales et
donc on ne tient pas compte du champ magn\'{e}tique.

D'apr\`{e}s les \S 6 et 8 de l'article d'Einstein, on a 
\begin{equation*}
\mathbf{E}_{\perp }^{\prime }=\gamma \mathbf{E}_{\perp }
\end{equation*}
\textbf{\ }Le mode de transformation relativiste n'\'{e}tant pas le m\^{e}me
pour les composantes transversales des deux vecteurs \textbf{E}\ et \textbf{A%
} puisque 
\begin{equation*}
\mathbf{A}_{\perp }^{\prime }=\mathbf{A}_{\perp }
\end{equation*}
nous disposons d'un crit\`{e}re math\'{e}matique de choix. Il est tout \`{a}
fait clair que selon ce crit\`{e}re ($d_{math\acute{e}matique}=A_{physique}$%
), on a: 
\begin{equation}
A=A^{\prime }=\text{ }A_{\perp }^{\prime }=A_{\perp }\qquad avec\qquad
A_{\parallel }^{\prime }\text{\ }=A_{\parallel }=0
\end{equation}
C'est bien le POTENTIEL VECTEUR $\mathbf{A}$ ($\mathbf{A}$') qui doit
\^{e}tre plac\'{e} sur chaque tangente aux cercles einsteiniens (cf. \textbf{%
Figure 1}). Bien entendu le champ \'{e}lectrique \textbf{E} l'est aussi,
mais l'\'{e}criture einsteinienne de l'onde plane \cite{1} impose \textbf{A
. k} = \textbf{A'. k' = 0} .

\begin{figure}
\begin{center}
\includegraphics[width=10cm]{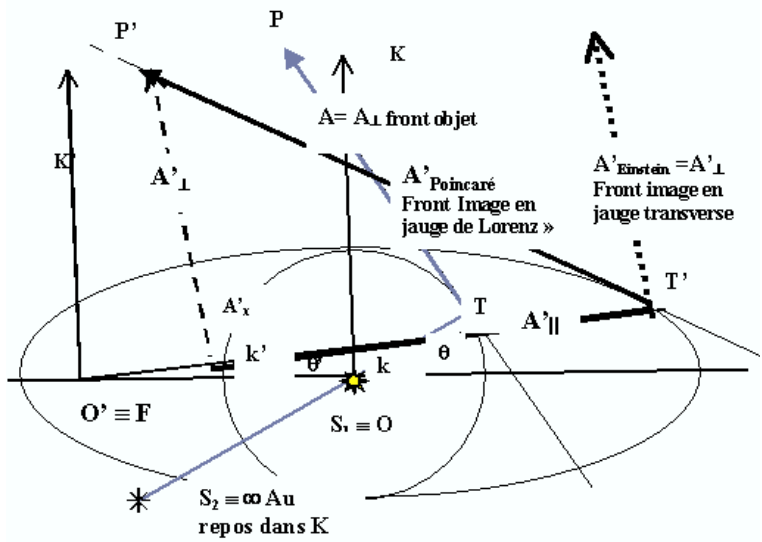}
\caption{Le vecteur TP du front objet se transforme par la TL en le vecteur T'P' sur le front image. La composante longitudinale
$A_{\parallel }^{\prime }$ du potentiel vecteur appara\^{i}t sur $k^{\prime }.$ Elle est annul\'{e}e par
Einstein.}
\end{center}
\end{figure}

\bigskip

Nous d\'{e}couvrons ainsi que le paragraphe 7 (phase de l'onde plane et
formule Doppler) de l'article d'Einstein de 1905 va bien au del\`{a} des
paragraphe 6 et 8 dans lesquels Einstein montre la covariance des
\'{e}quations de Maxwell, \ respectivement sans source ($\rho =0)$ et avec
source \cite{21}. Dans ces deux derniers paragraphes, Einstein s'en tient
strictement \`{a} la transformation relativiste du champ
\'{e}lectromagn\'{e}tique. Le choix implicite du potentiel pour l'onde plane
dans le paragraphe 7 ne change strictement rien au niveau des champs. Mais
par contre il est d'une importance cruciale pour la cin\'{e}matique
einsteinienne car \textit{physiquement} c'est bien le mode de
d\'{e}termination du potentiel vecteur \textbf{A} qui r\`{e}gle le
probl\`{e}me de la simultan\'{e}it\'{e} (et non pas du champ \'{e}lectrique 
\textbf{E}). Nous insistons une nouvelle fois sur le fait qu'Einstein n'a
jamais mentionn\'{e} les potentiels dans ces \'{e}crits. Cependant, la
traduction \textit{en termes des potentiels \'{e}lectromagn\'{e}tiques} de
l'\'{e}criture par Einstein des ondes planes impose une condition de jauge
compatible avec la double transversalit\'{e} et la double
simultan\'{e}it\'{e}.

Le potentiel vecteur $\mathbf{A}$ conserve sa norme $A$ qui est un invariant
implicite de la cin\'{e}matique einsteinienne: 
\begin{equation}
A^{\prime }=A
\end{equation}
Et en vertu m\^{e}me de l'invariance de la norme du quadrivecteur (du genre
spatial) potentiel, on a: 
\begin{equation}
V^{2}-A^{2}=V^{^{\prime }2}-A^{\prime 2}
\end{equation}
et donc n\'{e}cessairement:
\begin{equation}
V=V^{\prime }=0
\end{equation}
On voit ainsi que la repr\'{e}sentation einsteinienne ne revient pas
seulement \`{a} ignorer l'ellipse (et sa tangente) mais bien plut\^{o}t \`{a}
''l'annuler''(puisqu'elle engendre une composante longitudinale non-nulle :
cf. \textbf{Figure 1}). Il est clair que cela correspond implicitement \`{a}
la \textbf{jauge ''transverse''(}ou Coulomb compl\'{e}t\'{e}e\textbf{) :} 
\begin{equation}
\nabla .\mathbf{A}=\nabla .\mathbf{{A^{\prime }}}=0
\end{equation}
\`{a} savoir la condition de jauge de Coulomb accompagn\'{e}e de
''l'annulation du potentiel scalaire $V=V^{\prime }=0$\ \ \textit{pour les
ondes}''.

Les caract\'{e}ristiques en jauge ''transverse'' de la grandeur physique
vecteur potentiel ($A_{\parallel }^{\prime }$\ $=0)$ sont bien celles que
l'on attendait: la jauge ''transverse'' est directement reli\'{e}e \`{a} la
question de la simultan\'{e}it\'{e} absolue (propagation instantan\'{e}e du
potentiel scalaire). Nous avons vu que Maxwell lui-m\^{e}me avait
conserv\'{e} la jauge de Coulomb pour le potentiel scalaire (Laplacien) car
une propagation non-instantan\'{e}e de l'interaction coulombienne
(D'Alembertien) lui para\^{i}ssait impossible physiquement (violation
flagrante du principe dynamique newtonien de r\'{e}action qui suppose la
simutan\'{e}it\'{e} absolue). Notre analyse montre que la traduction en
termes des potentiels de la d\'{e}marche einsteinienne revient \`{a} adopter
la position de Maxwell sur ce point en ajoutant l'annulation du ''potentiel
des ondes'' pour d\'{e}finir ses fronts d'onde purement spatiaux dans les
deux syst\`{e}mes.

Remarquons que dans ce dernier cas de figure, la jauge de Lorenz \'{e}crite
comme une condition longitudinale gr\^{a}ce \`{a} la d\'{e}composition de
Helmholtz (voir \S\ 2.2) est trivialement v\'{e}rifi\'{e}e par annulation 
\textit{radicale} de toutes les grandeurs qui s'y trouvent: 
\begin{equation}
V=V^{\prime }=0\quad A_{\parallel }=A_{\parallel }^{\prime }=0
\end{equation}
\ 

Un quadrivecteur amput\'{e} de la 1\`{e}re et de la 4\`{e}me composante
n'est cependant plus un quadrivecteur. La jauge de Coulomb \'{e}tait donc
bien cach\'{e}e puisqu'elle \'{e}tait ''tapie'' au coeur m\^{e}me de la
cin\'{e}matique einsteinienne. On comprend d\`{e}s lors pourquoi le front
image plan einsteinien viole la relativit\'{e} de la simultan\'{e}it\'{e}
puisqu'il est fond\'{e} sur le choix d'une jauge r\'{e}put\'{e}e...\
non-relativiste.

Si on annule la grandeur qui se propage instantan\'{e}ment, plus rien ne
semble se propager instantan\'{e}ment. Il est \'{e}vident que ce choix
implicite sur les potentiels ne concerne pas que les ondes planes. Il ne
saurait \^{e}tre question de ''saucissoner'' la cin\'{e}matique
einsteinienne (en ''tranches coulombiennes'' et ''tranches
non-coulombiennes''). Les fronts plans einsteiniens sont tangents aux fronts
sph\'{e}riques einsteiniens. Ces derniers sont ins\'{e}parables de la
convention einsteinienne de synchronisation et des tiges rigides
einsteiniennes (et de sa d\'{e}finition de la contraction). C'est bien donc
toute la cin\'{e}matique d'Einstein qui est induite par un choix de jauge,
celle de la jauge transverse.

\subsection{La cin\'{e}matique de Poincar\'{e} et la condition de jauge de
Lorenz}

\bigskip Afin d'\'{e}tablir d\'{e}finitivement l'existence d'une structure
fine de la relativit\'{e} restreinte, nous devons maintenant prouver que la
cin\'{e}matique poincar\'{e}enne est parfaitement compatible avec la
transversalit\'{e} des ondes \'{e}lectromagn\'{e}tiques au niveau des champs
et qu'elle est fond\'{e}e sur le choix de la condition de jauge de Lorenz.
L'enjeu est \'{e}videmment crucial puisque la formule Doppler \cite{1}\
d\'{e}pendrait alors d'un choix de jauge, ce qui est non-orthodoxe. On peut
toujours poser dans le syst\`{e}me objet $\mathbf{A}_{\parallel }=V=0$ mais
nous devons montrer qu'avec une transformation conforme \`{a} la jauge
(relativiste) de Lorenz, on obtient $\mathbf{A}_{\parallel }\neq V\neq 0$
sans introduire une composante longitudinale E$_{\parallel }$ au niveau des
champs (ici champ electrique), ce qui invaliderait bien entendu la
cin\'{e}matique induite de l'ellipse.

Introduisons maintenant la d\'{e}composition de Helmholtz. Elle correspond
au fait d'\'{e}crire n'importe quel vecteur comme la somme d'un gradient
\`{a} rotationnel nul et d'un vecteur sol\'{e}no\"{i}dal.

Effectuons maintenant une d\'{e}composition du potentiel vecteur pour une
onde \'{e}lectromagn\'{e}tique plane \cite{10} :

\begin{equation}
\mathbf{A}=\mathbf{A}_{_{_{\parallel }}}+\mathbf{A}_{\perp }=\nabla g+\nabla
\times \mathbf{R}
\end{equation}
ou g est un scalaire et \textbf{R} un vecteur.

Comme $\nabla .\mathbf{A}_{\perp }=0$ la condition de Lorenz ne met en jeu
que la partie longitudinale de $\mathbf{A}$

\begin{equation}
\nabla .\mathbf{{A}_{_{\parallel }}}+\frac{1}{c^{2}}\partial _{t}V=0
\end{equation}

Au contraire, il est facile de voir que les champs s'expriment seulement en
fonction de $\mathbf{A}_{\perp }$ sous la forme

\begin{equation*}
\mathbf{E}=-\partial _{t}\mathbf{A}_{\perp }\qquad \qquad \mathbf{B}=\nabla
\times \mathbf{A}_{\perp }
\end{equation*}

Pour $\mathbf{B}$, cela r\'{e}sulte simplement de $\nabla \times \mathbf{A}%
_{\parallel }=0.$ Pour $\mathbf{E}$ c'est une cons\'{e}quence de la
contrainte de Lorenz (par transform\'{e}e de Fourier on a $V_{L}=c_{L}A_{x}$%
). Il est alors facile de voir que le terme en $-\partial _{t}A$ du champ
\'{e}lectrique est compens\'{e} par le gradient du potentiel scalaire $%
-gradV $.

La d\'{e}composition de Helmholtz permet de ''longitudinaliser'' la jauge de
Lorenz qui devient ainsi une condition sur la composante longitudinale du
potentiel vecteur et le potentiel scalaire. La transversalit\'{e} des ondes
\'{e}lectromagn\'{e}tiques au niveau des champs \'{e}lectromagn\'{e}tiques
n'impose pas la jauge de Coulomb. En effet, dans l'expression du champ
\'{e}lectrique longitudinal, les deux termes se compensent avec \cite{10} : 
\begin{equation}
V=cA_{x}
\end{equation}

Dans le raisonnement ci-dessus, il s'agit clairement d'une propagation
longitudinale o\`{u} $A_{x}=A_{\parallel }.$ Or la d\'{e}composition de
Helmholtz est d\'{e}finie sur la direction de propagation du front
ind\'{e}pendamment de la direction du mouvement de l'observateur. On doit
donc pouvoir obtenir une compensation quel que soit l'angle $\theta $ entre
la propagation (de l'onde) et le mouvement (de K' par rapport \`{a} K).
Autrement\ dit que se passe-t-il si $A_{x}$ ne correspond pas \`{a} une
composante longitudinale? Ecrivons la TL \`{a} deux dimensions spatiales
sous la forme suivante (\cite{18}, $c=1$), les deux sources
consid\'{e}r\'{e}es \'{e}tant au repos dans le syst\`{e}me K (le front plan
objet est dans K et le front plan image est dans K': cf. \textbf{Figure 1}): 
\begin{equation}
x^{\prime }=\gamma (x+\beta t)\qquad \qquad \qquad y^{\prime }=y\qquad
\qquad t^{\prime }=\gamma (t+\beta x)
\end{equation}

Projetons respectivement $\mathbf{A}$\ et $\mathbf{A}^{\prime }$ sur le
syst\`{e}me d'axe perpendiculaire Oxy dans K et sur Ox'y' dans K' et sur le
syst\`{e}me d'axes perpendiculaires form\'{e} par la direction de
propagation $\mathbf{k}$ dans K $(\mathbf{k}^{\prime }$ dans K'$)$ et la
perpendiculaire \`{a} cette direction:

\begin{equation*}
A^{\prime 2}=A_{x}^{\prime 2}+A_{y}^{\prime 2}=\gamma ^{2}A_{x}^{2}+A_{y}^{2}
\end{equation*}
\begin{equation}
A_{x}^{\prime }=\gamma A_{x}\qquad A_{y}^{\prime }=A_{y}
\end{equation}

Il est tr\`{e}s ais\'{e} de construire un quadrivecteur \cite{1} dont les
composantes se transforment comme (16). Il suffit d'annuler la quatri\`{e}me
composante du quadrivecteur dans K (ce qu'on peut toujours faire avec un
4-vecteur du genre espace)

\begin{equation*}
(A_{x},\qquad A_{y}\ ,\ \ \ \ \ 0,\qquad 0)
\end{equation*}

dont la norme est:

\begin{equation}
\left\| (A_{x},\qquad A_{y}\ ,\ \ \ \ \ 0,\qquad 0)\right\| =A_{\perp }^{2}
\end{equation}

La quatri\`{e}me composante est alors:

\begin{equation*}
A_{x}^{\prime }=\gamma (A_{x}+\beta 0)\qquad A_{y}^{\prime }=A_{y}\text{ \ \
\ \ \ \ }V^{\prime }=\gamma (0+\beta A_{x})
\end{equation*}

Le 4-vecteur s'\'{e}crit dans Oxy:

\begin{equation}
(A_{x}^{\prime },\qquad A_{y}^{\prime },\ \ \ \ \ \ \ 0,\qquad V^{\prime })
\end{equation}

dont la norme est

\begin{equation}
\left\| (A_{x}^{\prime },\qquad A_{y}^{\prime },\ \ \ \ \ \ \ 0,\qquad
V^{\prime })\right\| =A^{\prime 2}-V^{\prime 2}
\end{equation}

Les formules de transformation des composantes du quadrivecteur
s'\'{e}crivent:

\begin{equation}
A_{x}^{\prime }=\gamma A_{x}\qquad A_{y}^{\prime }=A_{y}\text{ \ \ \ \ }0%
\text{\ \ \ \ \ \ }V^{\prime }=\gamma \beta A_{x}
\end{equation}

\bigskip On voit ainsi que la compensation (14) ne fonctionne ni dans K ni
dans K'. C'est \'{e}videmment normal puisque dans le cas d'une propagation
non-longitudinale de l'onde plane, $A_{x}$ ($A_{x}^{\prime })$ n'est pas la
composante longitudinale de l'onde. D\'{e}signons cette composante
longitudinale sur la direction de propagation par $A_{\parallel }^{\prime }$
qu'il convient de calculer. Supposons que la d\'{e}composition de Helmholtz
s'\'{e}crive de la m\^{e}me mani\`{e}re dans le syst\`{e}me prim\'{e}.

\begin{equation}
\mathbf{A}^{\prime }\mathbf{=A}_{\parallel }^{\prime }\mathbf{+}\text{ }%
\mathbf{A}_{\perp }^{\prime }\mathbf{\qquad }\qquad A^{\prime
2}=A_{\parallel }^{\prime 2}+A_{\perp }^{\prime 2}
\end{equation}

autrement dit

\begin{equation*}
\gamma ^{2}A_{x}^{2}+A_{y}^{2}=A_{\parallel }^{\prime 2}+A^{2}\qquad \gamma
^{2}A_{x}^{2}=A_{\parallel }^{\prime 2}+A_{x}^{2}
\end{equation*}

Et donc 
\begin{equation}
A_{\parallel }^{\prime }=\gamma \beta A_{x}
\end{equation}

On voit ainsi clairement que la compensation se produit pour tout $\theta
\neq \theta ^{\prime }\neq 0$ car

\begin{equation}
V^{\prime }=A_{\parallel }^{\prime }=\gamma \beta A_{x}
\end{equation}

En calculant la norme du 4-vecteur 
\begin{equation}
\left\| (A_{x}^{\prime },\qquad A_{y}^{\prime },\ \ \ \ \ \ \ 0,\qquad
V^{\prime })\right\| =A^{\prime 2}-V^{\prime 2}=V^{^{\prime }2}-A_{\perp
}^{\prime 2}-A_{\parallel }^{\prime 2}=A_{\perp }^{\prime 2}
\end{equation}

La d\'{e}composition de Helmholtz prim\'{e}e conduit donc bien \`{a}
l'invariance (19 \& 24) de la norme $V^{^{\prime }2}-A_{\parallel }^{\prime
2}$ du quadrivecteur potentiel prim\'{e}. \textit{On a donc d\'{e}montr\'{e}
que la d\'{e}composition de Helmholtz ne d\'{e}pend pas de la direction de
propagation} et est donc invariante (12 \& 21) par la TL \cite{1}.

La transformation de LorenTz du quadrivecteur revient \`{a} travailler avec
la condition de jauge relativiste de Lorenz, \'{e}crite selon la
d\'{e}composition vectorielle de Helmholtz:

\begin{equation}
\nabla .\mathbf{{A}_{_{\parallel }}}+\frac{1}{c^{2}}\partial _{t}V=0\qquad
\nabla .\mathbf{{A}_{_{\parallel }}^{\prime }}+\frac{1}{c^{2}}\partial
_{t}V^{\prime }=0
\end{equation}
dans le premier syst\`{e}me par annulation $A_{_{\mid \mid }}=V=0$ mais pas
dans le syst\`{e}me image K' o\`{u} $A_{_{\mid \mid }}^{\prime }=V^{\prime
}\neq 0.$

Nous arrivons ainsi au but car nous devions montrer que l'ellipse n'est pas
en contradiction avec la transversalit\'{e} au niveau des champs. On voit
alors que la composante longitudinale du champ \'{e}lectrique

\begin{equation}
\mathbf{E}_{\parallel }=-\partial _{t}\mathbf{A}_{_{\mid \mid }}-\nabla
V\qquad \mathbf{E}_{\parallel }^{\prime }=-\partial _{t}\mathbf{A}_{_{\mid
\mid }}^{\prime }-\nabla V^{\prime }
\end{equation}
s'annule dans le syst\`{e}me prim\'{e} (image) par compensation.

Quelques jours avant sa mort, dans une s\'{e}rie de cours \`{a} l'Ecole
Sup\'{e}rieure des Postes et des T\'{e}l\'{e}graphes, Henri Poincar\'{e},
reviendra sur l'ellipse allong\'{e}e et sur son adoption d'une th\'{e}orie
\'{e}lectromagn\'{e}tique exprim\'{e}e en termes des potentiels. En effet on
peut voir sur la m\^{e}me page (46-47) de la publication posthume \cite{19}
par ses \'{e}l\`{e}ves-ing\'{e}nieurs t\'{e}l\'{e}graphistes (question de
synchronisation oblige!) d'une part l'ellipse (avec le r\^{o}le des
composantes longitudinales \cite{1}) et d'autre part la pr\'{e}\'{e}minence
de la th\'{e}orie de Riemann-Lorenz sur celle de Heaviside-Hertz (avec le
r\^{o}le des composantes longitudinales \cite{10}).

\section*{Conclusions}

Pourquoi Einstein est-il rest\'{e} silencieux sur les potentiels ? On sait
qu'il a appris principalement l'\'{e}lectromagn\'{e}tisme en lisant Hertz.
Par ailleurs, il est tr\`{e}s probable qu'il est eu acc\`{e}s au cours d'A.
F\"{o}ppl qui traitait de la th\'{e}orie de Maxwell en termes des
potentiels. En effet, plusieurs commentateurs y ont trouv\'{e} l'exemple du
mouvement relatif entre l'aimant et la spire qu'Einstein prit comme exemple
introductif de son \'{e}lectrodynamique des corps en mouvement. De plus, on
sait qu'Einstein lut le \guillemotleft\ Versuch \guillemotright\ de Lorentz
paru en 1895 donc il connaissait en 1905 les \'{e}quations faisant
intervenir les potentiels. N\'{e}anmoins, il adopta la formulation de Hertz
qui exclu les potentiels.

Nous voudrions sugg\'{e}rer une interpr\'{e}tation possible issue de sa
vision thermodynamique de la lumi\`{e}re \cite{23}. En effet, comme Pauli
l'a soulign\'{e}, il est probable qu'il y ait eu une influence des travaux
d'Einstein \`{a} propos du corps noir sur sa cin\'{e}matique relativiste. Le
d\'{e}nombrement des modes du champ dans la cavit\'{e} du corps noir fait
intervenir un facteur 2 dont l'origine physique correspond \`{a} la prise en
compte des deux polarisations transverses de la lumi\`{e}re. Einstein a-t-il 
\'{e}t\'{e} influenc\'{e} par la confirmation exp\'{e}rimentale de la loi de
Planck pour \'{e}carter la possibilit\'{e} de l'existence d'une propagation
longitudinale relative \`{a} une caract\'{e}ristique de l'onde \'{e}%
lectromagn\'{e}tique comme le potentiel vecteur ? Auquel cas, sa cin\'{e}%
matique ne pouvait \^{e}tre compatible qu'avec le caract\`{e}re transverse
thermodynamique de la lumi\`{e}re. Nous retrouvons ainsi la filiation
thermodynamique Planck-Einstein que l'un d'entre nous avait mis en \'{e}%
vidence \`{a} propos de la formulation de la relativit\'{e} restreinte \cite
{2}.

Nous avons donc montr\'{e} que la cin\'{e}matique relativiste de Poincar\'{e}%
, enti\`{e}rement fond\'{e}e sur une th\'{e}orie purement ondulatoire de la
lumi\`{e}re, est tout \`{a} fait compatible avec l'existence d'une
propagation longitudinale au niveau des potentiels telle que d\'{e}crite par
la formulation de Riemann-Lorenz. Il reste \`{a} examiner la mani\`{e}re de
quantifier le champ \'{e}lectromagn\'{e}tique dans le cadre de cette
nouvelle cin\'{e}matique de l'espace-temps...

\end{document}